# The Morphologic Properties of Magnetic networks over the Solar Cycle 23

Chong Huang[1], Yihua Yan[1], Yin Zhang[1], Baolin Tan[1] and Gang Li[2,1]

[1], *Key Laboratory of Solar Activity, National Astronomical Observatories of Chinese Academy of Sciences, Beijing, China, 100012*

[2], *Department of Physics and CSPAR, University of Alabama in Huntsville, AL, 35899, USA*

chuang@nao.cas.cn, yyh@nao.cas.cn

## ABSTRACT

The morphologic properties of the magnetic networks during Carrington Rotations (CR) 1955 to 2091 (from 1999 to 2010) have been analyzed by applying the watershed algorithm to magnetograms observed by the Michelson Doppler Interferometer (MDI) on board the Solar and Heliospheric Observatory (SOHO) spacecraft. We find that the average area of magnetic cells on the solar surface at lower latitudes (within $\pm 50°$) are smaller than those at higher latitudes (beyond $\pm 50°$). Statistical analysis of these data indicates that the magnetic networks are of fractal in nature, and the average fractal dimension is $D_f = 1.253 \pm 0.011$. We also find that both the fractal dimension and the size of the magnetic networks are anti-correlated with the sunspot area. This is perhaps because a strong magnetic field can suppress spatially modulated oscillation, compress the boundaries of network cells, leading to smoother cell boundaries. The fractal dimension of the cell deviates that predicted from an isobar of Kolmogorov $k^{-5/3}$ homogeneous turbulence.

*Subject headings:* techniques:image processing—methods: statistical—Sun: activity—Sun: surface magnetism

## 1. Introduction

Solar magnetic fields are generated by the motion of conductive plasma in highly turbulent convective layers. Their emergence manifests the magnetic networks at the solar photosphere. Magnetic network is always considered as the main component of small-scale magnetism. In 1960s, Leighton and Simon (Leighton et al. 1962; Leighton 1964; Simon & Leighton 1964) found that the magnetic networks are strongly related to supergranulations observationally. Although researches in magnetic networks and supergranulations have undergone tremendous progresses, the interaction between magnetic network/supergranulation and magnetic activity is still not well resolved (Meunier, Roudier & Tkaczuk 2007).

In modeling the diffusion process that drives the dynamic evolution of the networks, it is useful to know the length scales and topologies of the magnetic networks. One important question is how the cell sizes vary with the magnetic environment. Employing a steepest descent algorithm, Hagennaar et al. (1997) studied a two-day sequence of Ca II k filtergrams and found no significant correlations between the magnetic network area and the magnetic flux content of the network cells. On the other hand, Komm et al. (1995) applied a cross-correlation method to the magnetograms observed by the National Solar Observatory (NSO) vacuum telescope from 1978 to 1990 (495 day pairs), and found a characteristic scale of $L = 16.6 \pm 0.2 Mm$ which does vary much (within %3) with solar cycle. Komm et al. (1995) also noted that the scale size increases with latitude during solar maximum when active regions present, in agreement with earlier results by Wang (1988); Wang et al. (1996). On the contrary, Singh & Bappu (1981); DeRosa et al. (2004) found that the scale size is smaller around solar maxima, consistent with the report of Meunier, Roudier & Tkaczuk (2007), who suggested that the average supergranular size anti-correlates with the magnetic activity.

Recently, (Bruno et al. (2001); Borovsky (2008); Li (2008); Li, Lee & Parks (2008)) it has been suggested that current-sheet-like structures observed in-situ at 1 AU may be the boundaries of flux tubes that originate from the magnetic networks of the solar surface. Indeed Bruno et al. (2001) suggested that the sizes of these flux tubes, when tracing back to the solar surface, may correlate with the photospheric magnetic networks. Using Ulysses observation, Miao, Peng & Li (2011) obtained the distribution of the waiting time statistics of the current sheets. If these current sheets are relic structures from the magnetic network on the solar surface, then one can relate such waiting time statistics with the distribution of the magnetic network sizes.

Fractal analysis has been used to quantify the complexity of geometric structure of turbulence, and thereby can shed light on the underlying dynamics (Mandelbrot 1977). For instance, fractal dimension of solar surface magnetic networks can be related to the underlying turbulence power spectrum, from which one can gain insights about the fundamental magnetohydrodynamic processes that involve the origin of magnetic structure (e.g. the local dynamo theory by Lawrence et al. (1993)). Since the fractal dimension can reflect the variation of cell shape, we use the fractal dimension to investigate how the shape of magnetic network is influenced by the magnetic activity, and analyze the relationship between the distribution of fractal dimension and the activity level over the solar cycle 23 (from 1999 to 2009).

In order to confirm the relationship between the cell sizes and the magnetic environment, we use the watershed algorithm (Gonzalez & Woods 2002) to identify the magnetic networks for the data of MDI/SOHO over the solar cycle 23, and perform statistic analysis of the cell sizes and the fractal dimension as a function of the solar activity level. We also examine magnetic network cell areas as a function of the latitude.

This paper is organized as follows: the data and analysis are briefly described in Section 2. The main property of the magnetic network (including the fractal dimension and the size) and the influence due to variation of magnetic activity is presented in Section 3. We conclude in Section 4.

## 2. OBSERVATIONS and DATA REDUCTION

### 2.1. Observations

We analyze the synoptic charts obtained by the Michelson Doppler Interferometer (MDI) on board the Solar and Heliospheric Observatory (SOHO) spacecraft (Scherrer et al. 1995) during the Carrington Rotation (CR) 1955 to CR 2091 (from 1999-10-11 15:45:34 UT to 2010-01-03 11:55:44 UT). Synoptic charts have been reconstructed using recalibrated magnetogram data. The pixel size is about $2''$ (about 1.450 Mm). We select a region of CR synoptic charts that situates between $15°$ S- $15°$ N latitude in heliographic coordinates to measure the morphologic properties of the magnetic networks. We also use synoptic charts situated at $60°$ S - $60°$ N to examine how the cell areas vary with latitude. In order to reveal the magnetic networks, we focus on the magnitude of the magnetic field only (i.e. all magnetic field data are replaced with their absolute values). We further filter out active regions by setting an upper threshold of the magnetic field strength to be 200 G.

## 2.2. Data reduction

The magnetograms contain not only magnetic networks but also other small-scale phenomena. A part of such region is shown in the left panel of Figure 1. Here the magnetic elements outline the boundaries of the networks – forming the cell patterns. We perform spatial smoothing to extract the magnetic network from other small-scale phenomena. The smoothing is done by convolution with a Gaussian function of a full-width at half-maximum (FWHM) of $\lambda$ ranging from 3 to 9 pixels (4.35-13.05 Mm). The effect of the smoothing scale $\lambda$ on the distributions of cell sizes is examined.

Fig. 1.— Left: Magnetogram sub-image obtained on Carrington Rotation 2025. The region is 200×200 pixels, corresponds to a field of about $400''\times400''$. Right: Watershed lines superimposed (thin solid line) on the same sub-image.

Using the smoothed data, we then identify the magnetic networks using the watershed algorithm. The essence of the watershed method is to treat the magnetograms as a topographic map, with the value of the magnetic field magnitude as height of various catchment basins (which are magnetic cells). When a water-drop falls on a this 2D map, it will enter one of these catchment basins. A water-drop falling exactly on the watershed ridge line will be equally likely to collect in adjacent catchment basins. As water drops continue to fall, the watershed algorithm finds the catchment basins and ridge lines in the image (Gonzalez & Woods 2002). In our case of the magnetograms, the network cells play the role of the catchment basins, and the boundaries of the cells correspond to the ridge lines.

Once the cells identified, we measure their areas and obtain the cell scale size $L$ through $L = A^{1/2}$. From $L$, we obtain the cell size distribution. We also measure the perimeter $P$ of the cell. From $A$ and $L$, we obtain the fractal dimension of the magnetic network.

To better examine the average size of the cells, we perform a Fast Fourier Transform (FFT) to the synoptic charts from year 2000 to 2009. Figure 2 shows the average FFT power spectrum for the whole period of 2000-2009 (the solid line), the solar maximum period 2000-2003 (the dashed line) and the solar minimum period 2005-2009 (the dotted line). We can see from the figure that there is a clear peak around 20 Mm ($\sim$ 14 pixels). This is the average size of the magnetic network. Comparing to the the watershed algorithm, we find that a smooth scale of $\lambda = 3.7$ pixels (5.08 Mm) yields the same average size of the magnetic network.

For checking this standard smoothing scale, we analyze the chromospheric cell from Ca II K filtergram with Precision Solar Photometric Telescope (PSPT), compare with magnetic network to show the influence of smoothing. We apply the same smoothing scale(3.7pixel) to extract the chromospheric network from the same region, shown in Figure 3. We found that the extracted chromospheric networks are almost match the manual extracted ones exactly, the cell boundaries are correspond to the locations of apparent enhanced intensity. The Probability Distribution of two different grams are shown in Figure 4, we can find the distributions are also similar. From the above, we can draw the conclusion that the magnetic network is match the chromospheric network well with a smoothing scale of 3.7 pixels, so we consider that use the smoothing scale of $\lambda = 3.7$ pixels (5.08 Mm), is standard to extract the network.

Fig. 2.— The average FFT power spectrum of the magnetograms during 2000-2009 (the solid line), solar cycle maximum 2000-2003 (the dash line), and solar cycle minimum 2004-2009(the dot line).

Fig. 3.— Left: The Ca II k filtergram sub-image corresponding to the same region in figure.1. Right: Watershed lines superimposed (thin solid line) on the same sub-image.

In the following, we therefore use a smoothing scale of $\lambda = 3.7$ pixels (5.08 Mm) in the watershed method. The right panel of Figure 1 shows an example of the magnetic networks obtained using the watershed method (with $\lambda = 3.7$ pixels). With such figures, one can examine various statistical properties of the magnetic networks.

Fig. 4.— Left: The Probability Distribution of chromospheric cell from figure.3. Right: The Probability Distribution of magnetic cell from figure.1

## 3. Result

### 3.1. Cell sizes distribution over the solar cycle 23

At present, the relation between cell size and the activity level is still highly debatable. As the prior result, some authors (e.g.Meunier, Roudier & Rieutord (2008); Singh & Bappu (1981)) compared the cell size with sunspot number, suggested that cell size is anti-correlated with sunspot number. For examining this result, we use sunspot area to compare with cell size. On the other hand, the sunspot area used here is to indicate the global activity level and the phase of solar cycle. The sunspot area also reflects the magnetic activity, the left panel of Figure 5 shows that the sunspot area and the magnetic activity is highly correlated.

We investigate the cells over ten years on magnetogram observations, obtaining the change of the mean network cell size for each Carrington Rotation. Over 574,000 cells are identified and measured, the results are shown in right panel of Figure 5 where the relation between cell scale size $L$ and the sun spot area is apparent. The smallest cell scale size is about 19.32 Mm which occurred at CR 1965 in solar maximum (from 2000.07.10 12:56:38 UT to 2000.08.06 17:39:12 UT), and the largest cell scale size is about 21.41 Mm which occurred in CR 2042 in solar minimum (from 2006.04.10 21:50:40 UT to 2006.05.08 03:50:40 UT). The average cell scale size is 20.04 Mm during solar cycle maximum (2000-2003). In comparison, the average cell scale size during the solar cycle minimum (2005-2009) is 20.53 Mm, which is larger by 2.37% than that of solar cycle maximum. The cell sizes decrease towards higher activity level, and the mean cell size during the whole 10 years is $20.32 \pm 0.48$ Mm, varies less than $\pm 2.35\%$, the linear correlation coefficient is $-0.444$. The result indicate that a moderate negative correlation between the average cell scale size and the sunspot area. The sunspot area in each CR is calculated from the daily sunspot area database from US Air Force (USAF) and the US National Oceanic and Atmospheric Administration (NOAA). This inverse correlation shows that on average a cell's size increases as solar activity decreases, in agreement with earlier works Singh & Bappu (1981); Berilli et al. (1998); Meunier, Roudier & Tkaczuk (2007).

The variation of magnetic network size can be considered as the reflection of changes in the structure of supergranulations. A change in magnetic field strength in the convection zone can affect the cell size. Chandrasekhar suggested a 5% reduction of cell size will lead to an enhancement in field by 15% (Chandrasekhar 1961). Weiss et al. (1996) studied the effect of magnetic field on the pattern of convection. They argued that a weaker magnetic field can not dominate the convective motion, therefore spatial modulated oscillations will become more violent and irregular during solar minimum, leading to a larger network scale.

Fig. 5.— Left: The average magnetic field intensity (the empty diamond) during the CR1955-2091 from 15° S to 15° N versus spot area (the filled ball). Right: Mean cell size (the empty ball) during the CR1955-2091 from 15° S to 15° N versus spot area (the filled ball).

### 3.2. The latitude dependence of the cell areas

We next examine how the magnetic network areas vary with latitude during the solar minimum and solar maximum periods. For solar minimum, the lower left panel of Figure 6 shows the mean magnetic network area as a function of latitude (60° S - 60° N). All the areas at each latitude are calibrated as $S/cos(\theta_{lat})$, and they are normalized to the value at equator, $S_m$. From the figure we can see that the mean cell area decreases towards the high latitude, from equator to about $\pm 50°$, reaching a minimum at about $\pm 30° - 40°$. The area then increases towards higher latitudes ($\pm 50°$). The variation of cell areas is about 15% within latitudes $\pm 50°$. The latitudinal dependence of the cell area agrees with (Rimmele & Schröter 1989), who reported that the mean cell size at latitudes around $\pm 45°$ is about 10% smaller than that at the equator. They also noted a possible increase of the size towards higher latitudes.

Comparing to solar minimum, the average cell area during solar maximum has a more complicated latitudinal dependence. The lower right panel of Figure 6 shows the mean magnetic network area as a function of latitude (60° S - 60° N) for the solar maximum periods: 2000-2003. There appears to be an asymmetry between northern hemisphere (positive latitude) and southern hemisphere (negative latitude). In the southern hemisphere, the latitudinal dependence of the cell size is similar to that during the solar minimum period: the mean cell area first decreases towards the high latitude and then increases again. In the northern hemisphere, however, the average cell size, except in year 2003, seems to monotonically increase with latitude.

The magnetic activity during solar minimum and maximum are also shown in upper row of figure 6, respectively. We consider that in solar maximum, an asymmetry in magnetic activity (in the same way, the butterfly diagrams) mainly leads to north-south asymmetry in cell area, but the relation between them is more complicated in solar minimum, which is worthy of further study.

Fig. 6.— Upper left: the average magnetic field intensity as a function of latitude during solar minimum periods 2004-2009. Upper right: the average magnetic field intensity as a function of latitude during solar maximum periods 2000-2003. Lower left: the scaled average network area as a function of latitude during solar minimum periods 2004-2009. Lower right: the scaled average network area as a function of latitude during solar maximum periods 2000-2003. Note the areas have scaled to the network area on the equator.

### 3.3. Fractal Dimension distribution

A fractal analysis can quantify the irregularity of a magnetic network and reveal the intrinsic features of the turbulence on the solar surface. Here we apply a fractal analysis which follow Mandelbrot (1977); Roudier & Muller (1986); Nesme-Ribes et al. (1996) on the magnetic networks obtained from the above watershed algorithm. The fractal dimension $D_f$ is defined by the area-perimeter relationship: $P \propto A^{D_f/2}$, where $P$ is the perimeter and $A$ is the area. For smooth shapes such as circles and squares, $D_f = 1$. We calculate the fractal dimension of magnetic networks in two CR (CR 2080 in solar minimum and CR 1982 in solar maximum). We show the perimeter $P$ versus area $A$ in a log-log plot for the cells within 15° S to 15° N in Figure 7. From the linear fit to the log-log plot, we can obtain the fractals $D_f$s for these two CRs, which are 1.238 and 1.269, respectively.

Fig. 7.— The perimeter P versus area A in a log-log plot with the data from 15° S to 15° N, the black solid line is a linear fit. Left is for CR 1982, and right is for CR 2080.

Since the area of sunspots can be regarded as an indicator of the solar activity level, we can examine the relationship between the fractal dimensions of network cell and the solar activity level. Figure 8 plots the fractal dimension $D_f$ and the sunspot area for all CRs investigate in this paper. The average fractal dimension is $D_f = 1.253 \pm 0.011$ from CR1955 to CR2091. One can see from the figure that periods of weaker activities correspond to larger fractal dimension and periods of stronger activities correspond to smaller fractal dimension. The minimum $D_f$ is about 1.231, occurred in CR 1989 in solar maximum (from 2002.04.26 07:32:06 UT to 2002.05.23 12:59:29 UT) and the maximum $D_f$ is about 1.273, occurred in CR 2090 in solar minimum (from 2009.11.09 20:33:44 UT to 2009.12.07 04:03:59 UT). The correlation coefficient is -0.707. This indicates that the correlation between the fractal dimension of magnetic networks and the sunspot area is stronger than the correlation between the average scale size of cells and the sunspot area.

The Kolmogorov theory of homogeneous and isotropic turbulence predicts that $D_f = 1.66$ and $D_f = 1.33$ for a two dimensional cut of isotherm and isobar respectively (Mandelbrot 1977). Our value $D_f = 1.253 \pm 0.011$ is smaller than both the isothermal and the isobar predictions. This fact implies that the strong magnetic field may suppress spatially modulated oscillation, reduce the scale of cell, and compress the boundaries of them, make the boundaries smoother.

In the classic homogeneous turbulence theory, the variance of the velocity follow the relationship(Batchelor

Fig. 8.— The comparison between fractal dimensions (the empty square) and sunspot areas (the filled ball) during CR1955 - 2091 from 15° S to 15° N. Error bars are ±1 standard deviation of the mean.

& Goldstein 1953):

$$\overline{|\text{VarU}|^2} = 4 \int_0^\infty E_U(k)(1 - \frac{\sin kr}{kr}) \, dk, r \ll L \tag{1}$$

where $U$, $k$, $L$ represent the velocity, the wave number and the integral length scale, respectively. $E_U(k)$ is the energy spectra, it can be expressed by $E_U(k) \propto k^\alpha$. The variance of velocity is therefore,

$$\overline{|\text{Var}U|^2} \propto r^{-(\alpha+1)}, \tag{2}$$

and in Batchelor's result, the relationship between variance of pressure and variance of velocity is

$$\overline{|\text{Var}P|^2} = \rho^2 (\overline{|\text{Var}U|^2})^2, \tag{3}$$

Here, $\rho$ is the mass density. According to Mandelbrot (1977), the fractal dimension of iso-surface is

$$D_f = D_E - \frac{1}{2} V_p, \tag{4}$$

where $D_E$ is the Euclid dimension and $V_p$ is the exponent of the pressure variance given by the following,

$$\overline{|\text{VarP}|^2} \propto r^{V_p}, \tag{5}$$

From equations (1), (2) and (4), one finds $\alpha \sim -1.75$. This is arguably agree with the power index of magnetic spectrum which is measured by Abramenko et al. (2001), which is -1.3 and -1.7 in quiet region and active region, respectively. This is also to be compared with the result of Matthaeus et al. (1982), who obtained the exponent of the power spectrum of the magnetic energy in the solar wind to be $1.73 \pm 0.08$. Our result is consistent with the scenario that the solar wind is emanating from the boundaries of magnetic network cells (Donald et al. 1999). Since Close et al. (2005) pointed out that several open-flux bundles in each supergranular cell connect from the photosphere into the solar wind, our results therefore supports the idea that current sheets observed in the solar wind are adjacent boundaries of flux tubes which originate from the solar surface, as advocated by (Bruno et al. 2001; Borovsky 2008; Li 2008; Li, Lee & Parks 2008).

## 4. Conclusion

In this work, we used a watershed algorithm to identify the magnetic network cells from the MDI data. We then obtained various statistical and geometric properties of the magnetic networks. Based on these analysis, we found:

(1) The average cell scale size weakly anti-correlates with the solar activity (or the sunspot area).

(2) The fractal dimension of the magnetic networks is $D_f = 1.231 - 1.273$. This is arguably consistent with the result of supergranulation by Paniveni et al. (2009), who report that $D_f = 1.12$ for solar active region, and 1.25 for solar quiet region. However, these authors also noted that they have not enough samples to confirm their values of $D_f$ (they used merely 200 networks in their analysis). In comparison, our result is derived from a much larger sample (more than 574,000 networks during a 10-year span). As the average cell size, the fractal dimension of magnetic network cells also anti-correlates with the activity level, and the correlation is stronger.

(3) The derived exponent of the energy spectrum of magnetic networks is roughly consistent with the exponent of magnetic energy in the solar wind. This result is consistent with the scenario that the solar wind is emanating from the boundaries and boundary intersections of magnetic network cells, and therefore implies that current sheets observed in-situ could originate from the solar surface.

(4) The magnetic network area is also latitude dependent. Between $0-\pm 50°$, it decreases with increasing latitude during solar cycle minimum. At latitudes higher than $\pm 50°$, it increases with increasing latitude. Such a latitudinal dependence will put a strong constraint to a specific physical model of the magnetic networks. However, We also note that the average cell size seems to monotonically increase with latitude in the northern hemisphere during 2000-2002, which will be discuss at the follow-up.

The authors are grateful to the SOHO team and PSPT team for providing the data. The work are supported by NSFC Grant No. 10921303, 10903013, 11273030, MOST Grant No. 2011CB811401, the National Major Scientific Equipment R&D Project ZDYZ2009-3. GL's work at UAHuntsville is supported by NSF grants ATM-0847719 and AGS-0962658.